\documentclass[11pt,twoside]{article}
\usepackage{newpasp,epsf}
\def\edcomment#1{\iffalse\marginpar{\raggedright\sl#1\/}\else\relax\fi}
\reversemarginpar

\begin{document}
\title{Models of Polarized and Variable Radio Emission for 
IDV Source 0917+624}
\author{T. Beckert, A. Kraus, T. P. Krichbaum, A. Witzel,
J. A. Zensus }
\affil{Max-Planck-Institut f\"ur Radioastronomie, Auf dem H\"ugel 69,
53121 Bonn, Germany}

\begin{abstract}
We examine the power spectra of IDV and show the information, which is to be
gained by wavelet analysis of light curves of the quasar 0917+624. Results for
total and polarized flux at 11\,cm are shown. Both interstellar scattering
and intrinsic models have difficulties in explaining the 1 day period
variations. A theoretical model for the time averaged emission is presented,
which provides the basis for the analysis of possible variations.   
\end{abstract}

\section{Introduction}
Intraday variability (IDV) in extragalactic radio sources has been found
in sources like 0917+624 (Quirrenbach et al. 1989) with variations on
time scales of about 1\,day (fast IDV), showing amplitudes
of 5-25\% in total flux.
The brightness temperatures of these sources range from $10^{18}$\,K
to $10^{21}$\,K,
if the variability time scale measures the source size.  
More recently ultrafast IDV sources PKS 0405-385 (Kedziora-Chudczer et al.
1997) and J1819+3845 (Dennett-Thorpe \& de Bruyn 2000) showing
time scales of half an hour to a few hours and much larger amplitudes have been
found. While refractive interstellar scintillation (RISS) seems to be the
only successful explanation for ultrafast IDV sources, the problem
has not been solved for longer period IDVs.

We compared the results of
Fourier power spectra and a wavelet analysis of light curves.
Published data are used (e.g. Qian et al. 1991)
from a 1989 multi-frequency campaign with the VLA and the
100\,m Effelsberg antenna at 3.6, 6, 11 and 20\,cm of total and polarized flux.
The quasar 0917+624 has a determined redshift of 1.44 and shows superluminal
motion up to\footnote{We assume
a matter dominated universe with Hubble constant $H_0 = 100h$ km s$^{-1}$
Mpc$^{-1}$ and deceleration $q_0 = 0.5$.}
$h\beta_{app} \la 8$ (Standke et al. 1996).

\section{Comparing Fourier Power Spectra and Wavelet Analysis}
While Fourier power spectra emphasizes periodic behaviour with the
best frequency resolution possible, a wavelet analysis is able to show at
what time intervals a typical time scale dominates the variations in the
light curve. It allows to identify short periods with specific rise or decay
times of individual bursts and one can find phase and frequency shifts.
\begin{figure}
 \plotfiddle{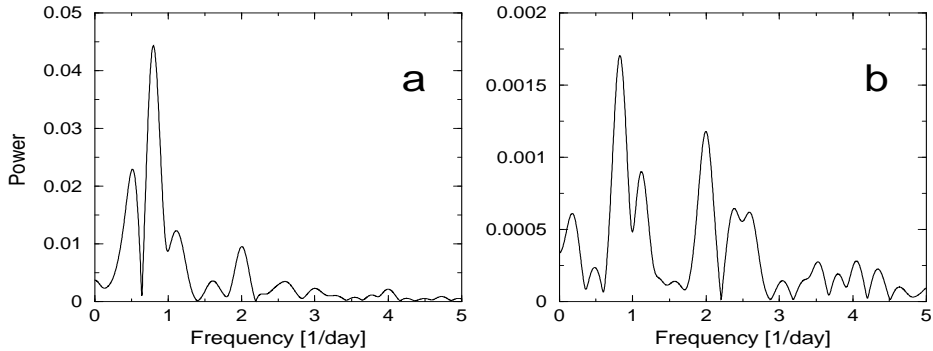}{3.9cm}{0}{64}{54}{-200}{-15}
 \caption{The CLEAN power spectrum of variations of 0917+624 at 11cm in
  total (a) and polarized flux (b). Data have been taken from the observing
  campaign in 1989. The peaks at $1.3$\,days and $0.5$\,days must be considered
  as significant in (a) and (b).}
\end{figure}
For the 0917+624 data, we use the CLEAN algorithm
for the Fourier transforms and a Morlet wavelet family (e.g.
Torrence \& Compo 1998). The spectra in Fig.\,1 and 2 show two characteristic
periods of 1.3 and 0.5\,days at 11\,cm consistently with both methods.
The  additional periods 
of 2 and 0.9\,days in Fig.\,1a are too weak and close to the 1.3\,day
period to be taken seriously. The
0.5\,day period is more pronounced in polarized than in total flux.
In addition the wavelet method shows, that this period in polarized flux P
varies in amplitude, but not in frequency during the 7\,days of
quasi-continuous observations.
\begin{figure}
 \plotfiddle{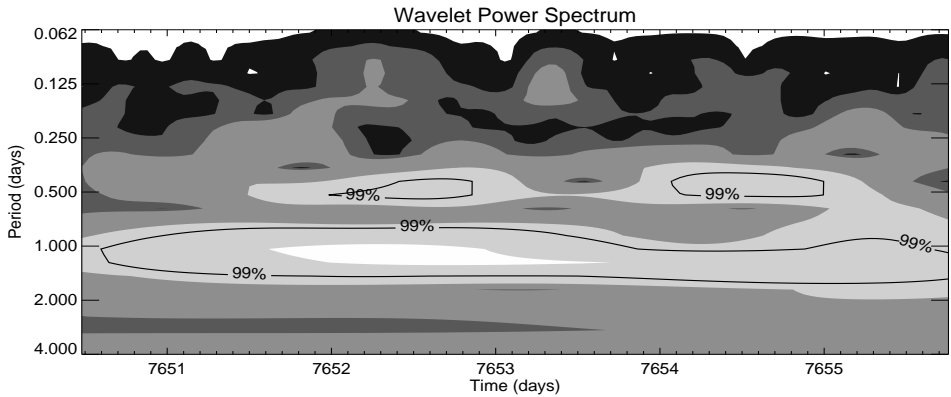}{4.6cm}{0}{54}{50}{-200}{-5}
\caption{Wavelet power spectrum for the same polarized flux data
          at 11\,cm as in Fig.\,1. The 99\% confidence level confirm the
          significance of the two periods of 1.3 and 0.5\,days.}
\end{figure}

\section{Interstellar Scintillation}
Annual variations of the variability time scale of J1819+3845 (Dennett-Thorpe,
this volume) provide a strong argument in favour of RISS as the sole
explanation for ultrafast variations, when interpreted as due
to the earth's motion with respect to the scattering medium. There is an
indication of similar behaviour in 0917+624 as shown by Kraus et al. (1999),
but changes in the characteristic time scale in the long-period IDV source
0716+714 have been found to
occur quasi instantaneously (Wagner et al., 1996). Radio IDV in this source
is correlated with variability at optical frequencies, which makes
this explanation very unlikely for 0716+714.
For 0917+624 we can approximate the variable component
with 10-15\% of Stokes I by a homogeneous ($\tau = 1$) source in the jet
moving with a Lorentz factor of $\Gamma = 10$ at an viewing angle of
$\vartheta = 0.5/\Gamma$ to the line of sight.  This leads to
the right apparent velocity $\beta_{app} = 8$ and a Doppler factor
$\delta =
16$. This compact component
can be steady over several years, if we assume energy equipartition and
a corresponding intrinsic brightness temperature\footnote{We assume a
spectral index of $p=2.5$ for the electron energy distribution.}
of $T_B = 5\cdot 10^{11} [\nu/ 1$GHz$]^{0.1}$.
From the 11\,cm data we derive a size of $4.3h^{-1}$ pc, which
corresponds to 1\,mas. For the scattering model of Rickett et al. (1995)
with a distance to the scattering screen of 200 pc and a velocity
of 50 km/s relative to the screen, the compact source has to be smaller than
0.1\,mas at 11\,cm, which is inconsistent with the equipartition
temperature model. Reducing the distance to the scattering screen to only
20\,pc as suggested by (Dennett-Thorpe \& de Bruyn, 2000) reduces the
discrepancy, but does not solve the problem.
Only a particle dominated source with a brightness temperature of
$5\cdot 10^{12}$\,K at 11\,cm is so small, that it resolves the problem,
but will violate the inverse Compton limit by $T_B = 8 T_{IC}$ and
predicts an unsteady source.

\section{Intrinsic variability}
A common feature of long period IDV sources like 0917+624 are the flat
spectra of their cores, which requires an inhomogeneous jet model.
The most promising model here is the propagation of a shock along the jet,
at which electrons are accelerated and radiate in the post-shock gas.
Variability is then caused either by density variations along the jet
or by directional changes of the gas motion in a helical relativistic jet.
In both cases the emission region must be small along the line of sight
compared with its transversal width. If variations are expected predominantly
along the jet, the viewing angle $\vartheta$ must be significantly smaller
than $1/\Gamma$, so that the line of sight is still at a small angle to the
jet direction in the jet rest frame.

Qian et al. (1991) argue that a shock can be naturally thin in
a relativistic jet. The post-shock gas will leave the shock with its sound
speed $\sim 0.6 c$ in the shock rest frame. This implies that the
shocked gas extents about
\begin{equation}
 \Delta R = \left(1-\sqrt{(\Gamma^2 - 4)/(\Gamma^2 -1)}\right)R
\end{equation}
along the jet, where $R$ is the position of the shock.
For a jet with an opening angle of $0.9\deg$ moving with
a Lorentz factor $\Gamma = 10$ the shock thickness and the jet width are
equal. Effective cooling of the shocked gas is then required
to keep the radiating layer of shocked gas sufficiently thin.

For modelling the radiative transfer of polarized synchrotron radiation (Jones
\& O'Dell, 1977) we use a conical jet model with adiabatic particle cooling.
The small degree of polarization is either due to a helical $B$-field structure
seen face-on, or due to turbulence on top of a mean
magnetic field as shown in Fig.\,3. To reproduce the averaged spectrum,
the shock is located $3\cdot 10^{20}$\,cm
from the tip of the cone with an opening angle of $2\deg$
in a region with a mean magnetic field of $0.13$\,mG and a particle density of
$N_e = 5\cdot 10^{-5}$\,cm$^{-3}$. The low energy cut-off of the electron
distribution is at $\gamma = 15$ and determines the degree of internal
Faraday rotation and conversion from linear to circular polarization.
Both effects can be relevant in the region of interest,which extents down to
$1\cdot 10^{19}$\,cm.
 \begin{figure}
\begin{minipage}[t]{7cm}
 \plotfiddle{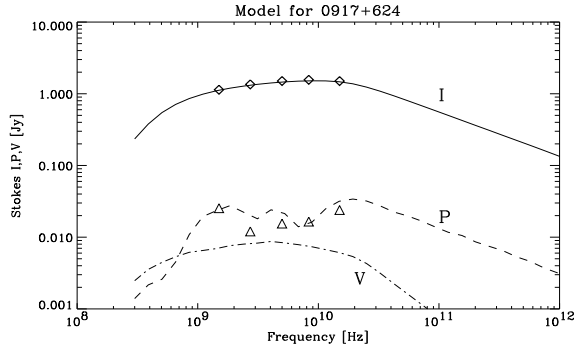}{4.2cm}{0}{55}{53}{-100}{0}
\end{minipage}
\begin{minipage}[b]{7cm}
 \caption{Averaged spectrum of the 1989 observing campaign for total
  (Stokes I: diamonds) and polarized flux (P: triangles). The model described
  in the text is shown as solid line for I, dashed for P, and dash-dotted
  for V. IDV in P is larger than the
  discrepancy of the modelled P and the averaged measurements.
}
\end{minipage}
\end{figure}
The radiating gas is assumed to move with  $\Gamma=10$ at an angle of
$\vartheta = 0.2/\Gamma$.

The model for the averaged emission can serve as a testbed for variability
caused either by RISS or by thin shocks in the jet. It is not possible to
determine the source of IDV in 0917+624 from present data.

\end{document}